\documentclass[aps,pra,twocolumn,amsmath,amssymb,nofootinbib,showpacs,superscriptaddress, showkeys]{revtex4-1}
\usepackage[pdftex]{graphicx}
\usepackage{amsmath}
\usepackage{amssymb}

\usepackage{hyperref}
\usepackage{cleveref}
\usepackage[svgnames]{xcolor}
\hypersetup{hidelinks,colorlinks=true,allcolors=DarkBlue}

\graphicspath{ {./images/} }
\newcommand{\tr}[1]{\textrm{#1}}

\begin{document}	
	
\title{Error estimation at the information reconciliation stage of quantum key distribution}
\author{E.O.~Kiktenko}
\affiliation{Russian Quantum Center, Skolkovo, Moscow, Russia 143025}
\affiliation{Steklov Mathematical Institute of Russian Academy of Sciences, Moscow, Russia 119991}
\affiliation{Geoelectromagnetic Research Centre of Schmidt Institute of Physics of the Earth, Russian Academy of Sciences, Troitsk, Moscow, 142190 Russia}
\author{A.O.~Malyshev}
\affiliation{Russian Quantum Center, Skolkovo, Moscow, Russia 143025}
\affiliation{Moscow Institute of Physics and Technology, Dolgoprudny, Moscow Region, Russia 141700}
\affiliation{Institute for Information Transmission Problems of the Russian Academy of Sciences,  Moscow 127051, Russia}
\author{A.A.~Bozhedarov}
\affiliation{Russian Quantum Center, Skolkovo, Moscow, Russia 143025}
\affiliation{Skolkovo Institute of Science and Technology, Moscow, Russia 121205}
\author{N.O.~Pozhar}
\affiliation{Russian Quantum Center, Skolkovo, Moscow, Russia 143025}
\author{M.N.~Anufriev}
\affiliation{Russian Quantum Center, Skolkovo, Moscow, Russia 143025}
\author{A.K.~Fedorov}	
\affiliation{Russian Quantum Center, Skolkovo, Moscow, Russia 143025}
	
\begin{abstract}
	Quantum key distribution (QKD) offers a practical solution for secure communication between two distinct parties via a quantum channel and an authentic public channel. 
	In this work, we consider different approaches to the quantum bit error rate (QBER) estimation at the information reconciliation stage of the post-processing procedure.  
	For reconciliation schemes using LDPC codes we develop a novel syndrome-based QBER estimation algorithm. 
	The suggested algorithm is suitable for irregular LDPC-codes, and takes into account punctured and shortened bits. 
	With testing our approach in the real QKD setup, we show that an approach combining the proposed algorithm with conventional QBER estimation techniques allows improving accuracy of the QBER estimation.	
\end{abstract}

\maketitle

\section{Introduction}\label{sec:Intro}
	
	Quantum computing possesses a threat on currently used information security protocols based on asymmetric cryptography~\cite{Shor1997}.
	A possible solution for establishing secure data transmission in the post-quantum era is to use the QKD technology~\cite{Gisin2002}. 
	This technology has attracted an enormous amount of interest~\cite{Gisin2002,Lo2015,Lo2016} last decades, and commercial QKD systems are now widely available on the market~\cite{Lo2016}. 
	
	Unlike conventional cryptographic tools, the security of QKD is based on the laws of quantum physics. 
	In particular, QKD protocols such as the seminal BB84~\cite{BB84} use individual quantum objects (e.g., photons) as information carriers. 
	However, classical communication and post-processing procedures are also required for QKD systems. 
	Due to noise in the quantum channel, after a quantum key establishment phase legitimate users have sifted keys that are weakly correlated and partially secure~\cite{Gisin2002}. 
	In order to correct down this error rate to the standard level, industrial QKD systems use information reconciliation procedures.
	
	The primary idea of information reconciliation is to remove discrepancies between keys of parties by disclosing some information over the authentic channel. 
	Further, disclosed bits of information shall be removed from the key within the privacy amplification procedure. 
	The less information is disclosed, the higher \emph{efficiency} a scheme demonstrates. 
	The efficiency and the number of additional communication rounds taken by scheme influence the rate of secret key generation.\,Thus, these are one of the main performance indicators of QKD setups.
	
	Among state-of-the-art reconciliation schemes one can distinguish protocols exploiting LDPC codes~\cite{Gallager1962,MacKay1999,Shokrollahi2004} to correct errors.
	They adopt an altered version of the conventional syndrome decoding algorithm to correct errors between blocks of the sifted keys.
	Significant improvements of the information reconciliation procedure with the use of LDPC codes have been suggested~\cite{Elkouss2010, Elkouss2011, LDPC_blind, Kiktenko2016, Kronberg2017, Kiktenko2017, Kiktenko2018}. 
	The latest development, called symmetric blind reconciliation protocol, is based on introducing symmetry in operations of parties, and the consideration of results of unsuccessful belief-propagation decodings~\cite{Kiktenko2017}.
	Consequently, this method allows to increase the procedure efficiency significantly and to reduce its interactivity, which leads to an increase in the secure key generation rate.
	
	We note that LDPC-based reconciliation schemes use an estimation of the QBER as one of the decoding algorithm parameters. 
	Due to that, inaccurate estimation of QBER can either result in the decline of scheme efficiency or lead to a higher number of additional communication rounds. 
	A common approach is to obtain QBER from the previous stages of the QKD workflow or to employ some default value of QBER typical for current QKD setup~\cite{Kiktenko2016}. 
	However, this lowers key generation rate since parties disclose redundant bits of the sifted keys to estimate the QBER.
	
	A distinct approach has been considered in Ref. \cite{qber-estimator}. 
	It is based on the use of syndromes of both parties to obtain on-the-fly estimation of the QBER for each block of the sifted key. 
	Unfortunately, an estimator considered in Ref.~\cite{qber-estimator} is suitable only for regular LDPC-codes and it does not take into account neither punctured nor shortened bits. 
	Meanwhile, shortening and puncturing techniques are important tools for the fine-tuning of LDPC-based schemes.
	
	In this paper we propose a novel on-the-fly QBER estimation approach providing the estimation of the QBER with the use of syndromes of parties both for irregular LDPC codes and in the presence of punctured or shortened bits. 
	In addition, we consider employing an a priori QBER distribution to increase accuracy of the estimation as well as combining this approach with the QBER estimation based on previous post-processing rounds.
	
	The paper is organized as follows. 
	In Section~\ref{sec:LDPC_scheme}, we explain basics of the LDPC-based information reconciliation. 
	In Section~\ref{sec:QBER_estimator}, we present the novel on-the-fly QBER estimation algorithm, which uses syndromes of the parties and an a priori QBER distribution. 
	Section~\ref{sec:Experimental} provides the results of the suggested algorithm trial in a real QKD setup.
	Finally, in Section~\ref{sec:Conclusion} we discuss our results and conclude the paper.
	
	\section{Information reconciliation with LDPC codes}\label{sec:LDPC_scheme}
	
	Here we consider a general scheme of the information reconciliation based on LDPC codes.
	We assume that Alice and Bob posses sifted keys $k_A$ and $k_B$ correspondingly.
	These keys are random bit strings of equal length.
	An $i$th bit of Alice's key $k_A[i]$ equals Bob's $i$th bit $k_B[i]$ with probability $1-q$ and differs otherwise.
	The value of $q$ (QBER) indicates how much information was intercepted during the transferring over a quantum channel.
	The aim of information reconciliation is to remove discrepancies between the sifted keys and make them identical on both sides.
	
	A common approach is to use a public authenticated channel for information reconciliation.
	However, we assume that all messages transferred through the public channel are possibly intercepted by an eavesdropper (Eve).
	Thus, parties shall also minimize Eve's knowledge about the sifted keys leaked during information reconciliation.
	
	One of the possible ways for removing discrepancies with a public discussion is to employ LDPC codes, which are linear error-correcting codes. 
	A particular LDPC code can be defined with the sparse binary $m\times n$ matrix ${\bf H}$ (with $m<n$) called parity-check matrix.
	
	A straightforward way of using an LDPC code is as follows.
	Alice computes her syndrome $s_A:={\bf H}k_A~({\rm mod}~2)$ and sends it to Bob (here the sifted key $k_A$ is considered as a column vector of length $n$).
	Then, Bob tries to find a key $\widehat{k}_A$, which is closest to ${k}_{B}$ such that ${\bf H}\widehat{k}_A~({\rm mod}~2)=s_A$.
	For that purpose he may use different approximate iterative algorithms of syndrome decoding such as sum-product~\cite{Decodoptim}, min-sum~\cite{minsum}, scaled min-sum~\cite{scaledminsum}, and others.
	Finally, after the decoding parties can use $\epsilon$-almost universal$_2$ ($\epsilon$-AU$_2$) hash functions to check whether $k_A$ and $\widehat{k}_A$ are equal up to small error probability $\epsilon$~\cite{Kiktenko2018}.
	
	Nevertheless, not an arbitrary LDPC code may be used for information reconciliation. 
	In paper~\cite{SlWolf} it has been demonstrated that to correct errors successfully (that is to obtain $\widehat{k}_A=k_A$) the following condition must be hold:
	\begin{equation}\label{eq:cond}
	f=\frac{m}{n h(q)}>1,
	\end{equation}
	where $f$ is called the efficiency of information reconciliation and $h(q)=-q\log_2q-(1-q)\log_2(1-q)$ is the binary entropy function.
	We note that $m$ is the length of a syndrome and it equals the amount of information leaked during the discussion.
	Thus, an effective information reconciliation protocol must both fulfil the condition given by Eq.~(\ref{eq:cond}) and disclose as little information as it is possible (i.e. $f$ shall not be too large).
	Actually, it turns out that $f=1.1\ldots 1.3$ is usually enough for successful error correction (see e.g. Ref.~\cite{LDPC_blind}).
	
	To perform fine tuning of the efficiency $f$, shortening and puncturing techniques can be employed~\cite{Elkouss2010}.
	The idea is to calculate a syndrome not from a block of the sifted key $k_A$ but construct an ``extendeded key'' $k_A^{\rm ext}$.
	In this case, $n_k$ bits are taken from the sifted key $k_A$, $n_s$ bits are fixed (shortened), and $n_p$ bits are initialized with true random values (punctured).
	The total number of bits $n_k+n_s+n_p$ equals the number of rows $n$ in the parity-check matrix ${\bf H}$.
	Bob performs the same operations to obtain an extended key $k_B^{\rm ext}$.
	Note that positions of key, shortened, and punctured bits inside extended keys come from pseudo-random number generators initialized with the same seed at Alice's and Bob's side.
	At the same time, values of punctured bits come from different truly random generators since they must be independent.
	
	In the presence of shortened and punctured bits, the expression of efficiency $f$ has the following from:
	\begin{equation}
	f = \frac{m-n_p}{(n-n_s-n_p)h(q)},
	\end{equation}
	as truly random punctured bits decrease the leakage of the information about sifted key by $n_p$ bits.
	Thus, variation of $n_s$ and $n_p$ (the value $n_s+n_p \equiv n_d$ is typically fixed) allows performing fine tuning of the efficiency $f$.
	
	Another way to improve the efficiency is to use additional communication rounds with extra bits from sifted keys disclosed.
	This approach is employed in blind~\cite{LDPC_blind} and symmetric blind~\cite{Kiktenko2017} information reconciliation protocols, and it allows improving performance of the information reconciliation procedure significantly.
	In this case the efficiency may be written as follows: 
	\begin{equation}
	f = \frac{m-n_p+n_{\rm add}}{(n-n_s-n_p)h(q)},
	\end{equation}
	where $n_{\rm add}$ is the number of bits disclosed in additional rounds.
	
	However, it is hard to achieve a desired efficiency since the actual number of discrepancies between $k_A$ and $k_B$ (or equivalently QBER) is unknown before the information reconciliation stage takes place. 
	To estimate $q$ Alice and Bob may employ the following techniques: 
	(i) disclose some sifted key bits at random positions over the public channel prior to information reconciliation and estimate $q$ by the number of discrepancies at that positions; 
	(ii) use some typical level of the QBER as an estimation; 
	(iii) use the value of the QBER from previous (already corrected) block of the sifted key; 
	(iv) make an estimation of the QBER using information from syndrome.  
	The method (i) has an obvious drawback since all the disclosed bits have to be discarded from the processed keys, while the method (ii) can be applied only in the case of very stable quantum channels.
	In the following section we consider methods (iii) and (iv).
	
	\section{Estimation of the QBER using syndromes} \label{sec:QBER_estimator}
	
	Accuracy of the QBER estimation strongly affects the efficiency of the information reconciliation. 
	For example, if QBER is overestimated, parties will use low-rate codes and the efficiency of reconciliation will be lower than possible. 
	In contrast, if QBER is underestimated, number of disclosed bits per round may be less than theoretical bound and there will be lots of additional communication rounds required. 
	In this section we describe a novel algorithm which uses information from syndrome to obtain more accurate QBER estimation.
	
	Let us consider a relative syndrome $\Delta s := s_A+s_B~({\rm mod}~2)$, where $s_B:={\bf H}k_B~({\rm mod}~2)$ is a Bob's syndrome.
	For now we assume that neither shortened nor punctured bits are used.
	According Ref.~\cite{qber-estimator}, for a regular LDPC code, that is a code with the parity-check matrix having the same numbers of unities in each row and each column, the bits of $\Delta s$ turn out to be i.i.d. 
	Bernoulli random variables with the following probability being equal to one:
	\begin{equation}
	p(q, d_c) := \Pr (\Delta s[i] = 1)= \sum \limits_{\substack{ j=1 \\  j \bmod 2 = 1}}^{d_c} \binom{d_c}{j} q^j  (1 - q) ^ {(d_c -j)}
	\end{equation}
	where $i=1,2,\ldots,m$, and $d_c$ is the number of unities in each row of the parity check matrix.
	To estimate the QBER $q$ from $\Delta s$ one can calculate maximum likelihood (ML) estimation of $p$ as follows: 
	\begin{equation}
	p_{\tr{est}}(\Delta s) := m^{-1} \sum_{i=1}^{m} \Delta s[i], 
	\end{equation}
	and then use it to obtain the ML estimation of $q$ according to the formulae:
	\begin{equation}\label{equ:simple_qber_ml_est}
	q_{\textrm{est}} (\Delta s) = - \frac{\left(1 - 2 \cdot p_{\tr{est}}(\Delta s)\right)^{\frac{1}{d_c}} - 1}{2}.
	\end{equation}
	We note that this estimator is suitable only for regular LDPC codes and it takes into account neither punctured nor shortened bits. 
	Below we consider an advanced version of the estimator which deals with both of these drawbacks.
	
	First, let us note that even for irregular LDPC codes all syndrome bits are still Bernoulli random variables. 
	However, in contrast to the case of regular code they have different crossover probabilities. 
	In particular, for $i$th bit of syndrome the following holds: 
	$\Pr (\Delta s[i] = 1)= p(q, d_c^{(i)})$, where $d_c^{(i)}$ is a number of unities in the $i$th row of the parity-check matrix.
	Therefore, we need to introduce a likelihood function of the QBER value $q$ for the relative syndrome $\Delta s$ in the following form:
	\begin{equation}\label{equ:syndrome_likelihood}
	\begin{aligned}
	L(q|\Delta s) = \prod_{\substack{i=1 \\ \Delta s[i] = 1}}^{m} p(q, d_c^{(i)}) \prod_{\substack{j=1 \\ \Delta s[j] = 0}}^{m} \left(1 - p(q, d_c^{(j)}) \right) = \\ =\prod_{i=1}^{m}\left[1-\Delta s[i]+(2\Delta s[i]-1)p(q,d_c^{(i)})\right].
	\end{aligned}
	\end{equation}
	Hence, the maximum likelihood estimation of QBER in the case of irregular LDPC code can be calculated as follows: 
	\begin{equation}
	q_{\rm est} = \arg \max \limits_{q \in [0, 0.5]} L(q|\Delta s).
	\end{equation}
	
	Second, we consider how punctured and shortened bits affect the bits of syndrome.
	Let us denote positions of shortened and punctured bits as $\mathcal{S}$ and $\mathcal{P}$ correspondingly.
	Remember, that we have $x_{A}^{\rm ext}[i]=x_{B}^{\rm ext}[i]$ for $i\in \mathcal{S}$, and two independent uniformly distributed random bits $x_{A}^{\rm ext}[i]$ and $x_{B}^{\rm ext}[i]$ for $i\in\mathcal{P}$.
	Let us also introduce position sets $\mathcal{A}_i=\{j : \mathbf{H}_{i,j}=1\}$, $i=1,2,\ldots m$, which are defined as sets of unity positions for each row of the parity-check matrix ${\bf H}$.
	
	Consider a particular position $i\in\{1,2,\ldots m\}$ in the relative syndrome $\Delta s$.
	One can see, that if $\mathcal{A}_i \cap \mathcal{P} \neq \emptyset$, then the value $\Delta{S[i]}$ turns out to be uniformly distributed random bit, and could not be used for an estimation of the QBER.
	Then, if $\mathcal{A}_i \cap \mathcal{P} = \emptyset$ we need to calculate a number of key bit positions inside $\mathcal{A}_i$.
	It is given by $\widetilde{d}_c^{(i)}:=d_c - |\mathcal{A}_i \cap \mathcal{S}|$, where $|\cdot|$ denotes cardinality of the set.
	Finally, in the presence of shortened and punctured bits, the likelihood function~(\ref{equ:syndrome_likelihood}) shall be rewritten in the following form:
	\begin{equation}\label{equ:syndrome_likelihood_adv}
	L(q|\Delta s) = \prod_{\substack{i=1 \\ \mathcal{A}_i \cap \mathcal{P}=\emptyset}}^{m}\left[1-\Delta s[i]+(2\Delta s[i]-1)p(q,\widetilde{d}_c^{(i)})\right].
	\end{equation}
	
	However, there are two more implementation peculiarities of QBER estimator to mention. 
	First, we propose to multiply likelihood \eqref{equ:syndrome_likelihood_adv} by the ``window'' function which takes into account typical values of QBER observed in experiments. 
	This allows avoiding outliers in the QBER estimation. 
	In particular, we use a combination of two sigmoid functions:
	\begin{equation}\label{equ:sigmoid}
	L_0(q) = \frac{1}{1 + e^{- \alpha_1  (q - q_{\tr{min}})}} \cdot \frac{1}{1 + e^{- \alpha_2  ( q_{\tr{max}} - q)}},
	\end{equation}
	where $\alpha_1$, $\alpha_2$, $q_{\tr{min}}$, and $q_{\tr{max}}$ are tunable parameters.
	
	Second, we propose to use log-domain for likelihood in the QBER estimation since it improves numerical stability of the method. 
	Thus, the function for ML estimation of QBER takes the form $\ell (q|\Delta s) = \log{L(q|\Delta s)} + \log{L_0(q).}$
	In order to find $q$ maximizing $\ell (q|\Delta s)$, we use standard optimization techniques.
	In particular, we apply Brent's method \cite{Brent_method} to the $(0.0, 0.5)$ interval.
	
	\section{Applications to industrial QKD systems}\label{sec:Experimental}
	\begin{figure}
		\centering
		\includegraphics[width=\linewidth]{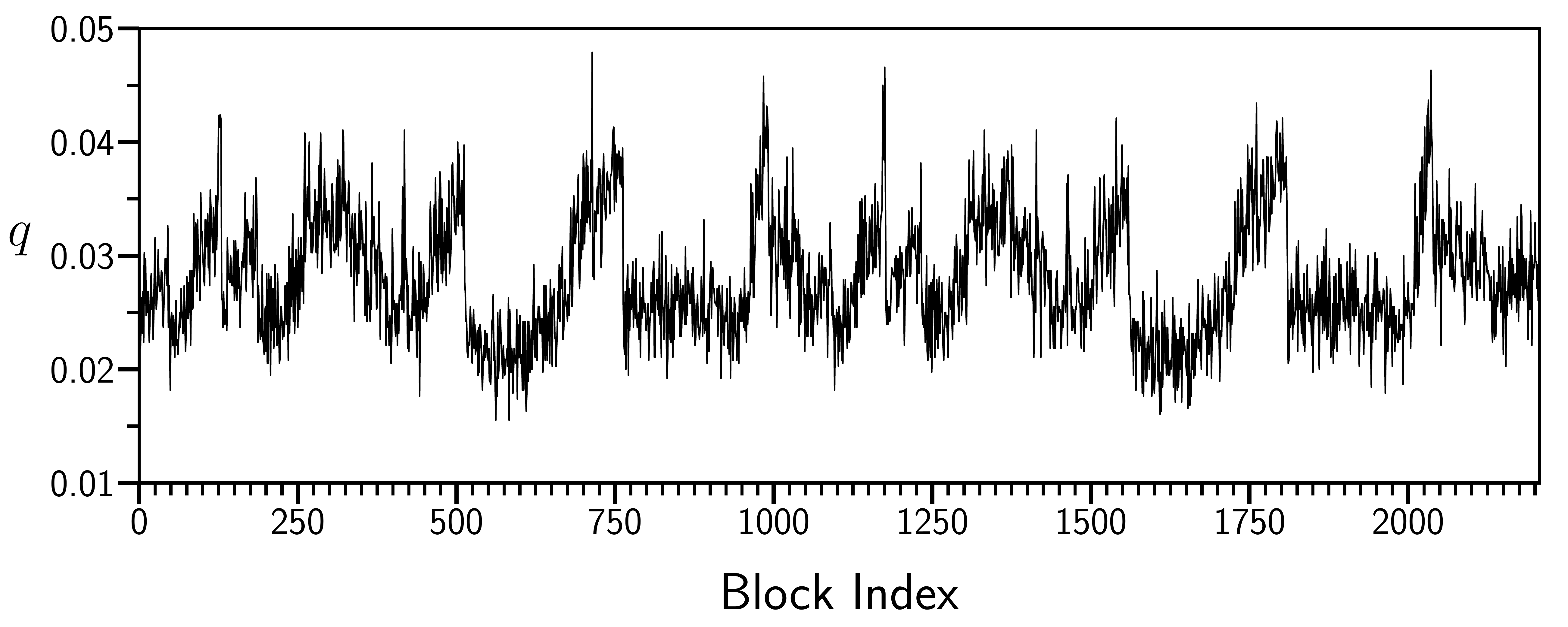}
		\caption{Experimental values of the QBER $q$ for 8Mbit sifted keys are shown as function of the key block index.
			Each point corresponds to a block of 3800 bits.}
		\label{fig:qber_to_time}
	\end{figure}
	
	\begin{figure*}
		\centering
		\includegraphics[height=0.35\linewidth]{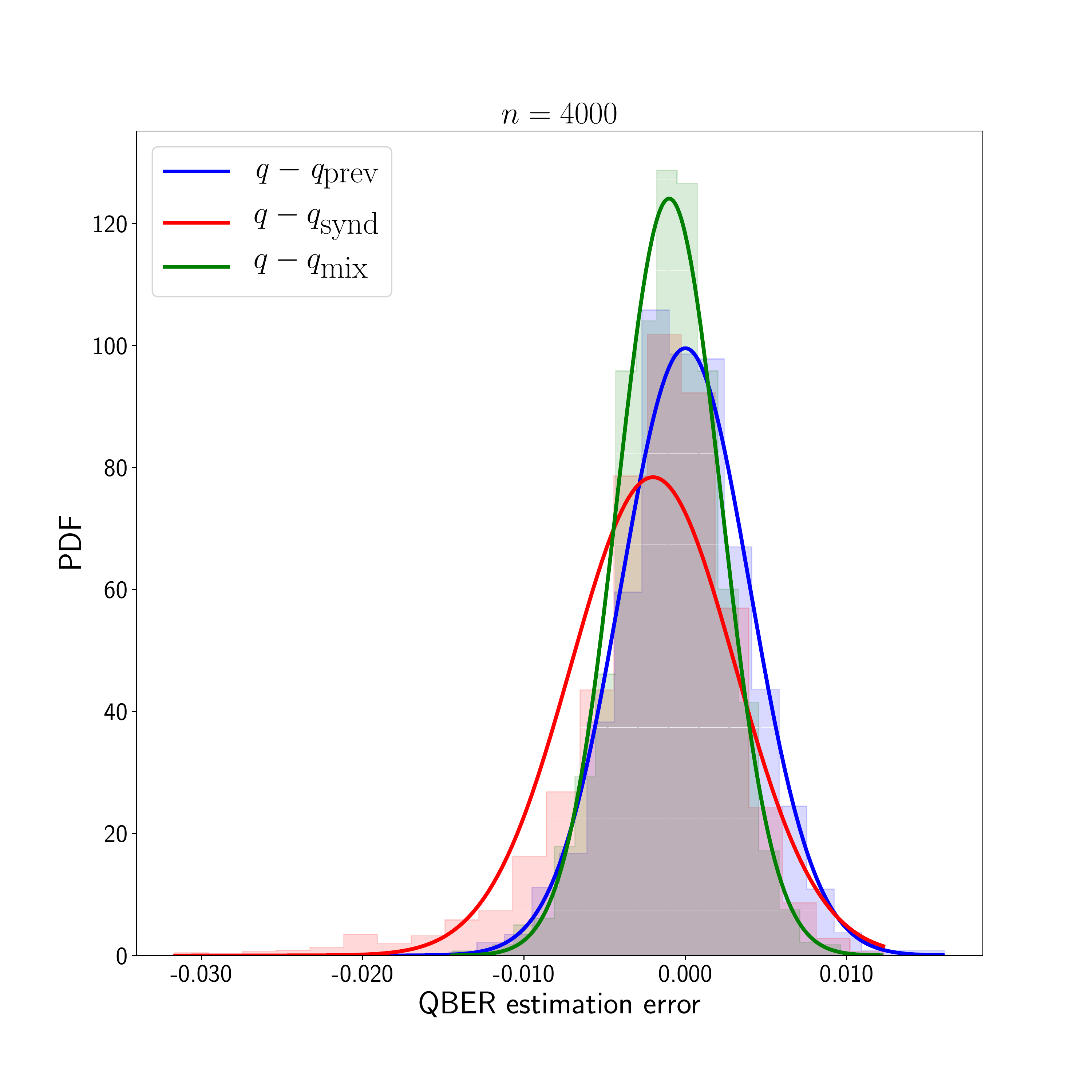} 
		\includegraphics[height=0.35\linewidth]{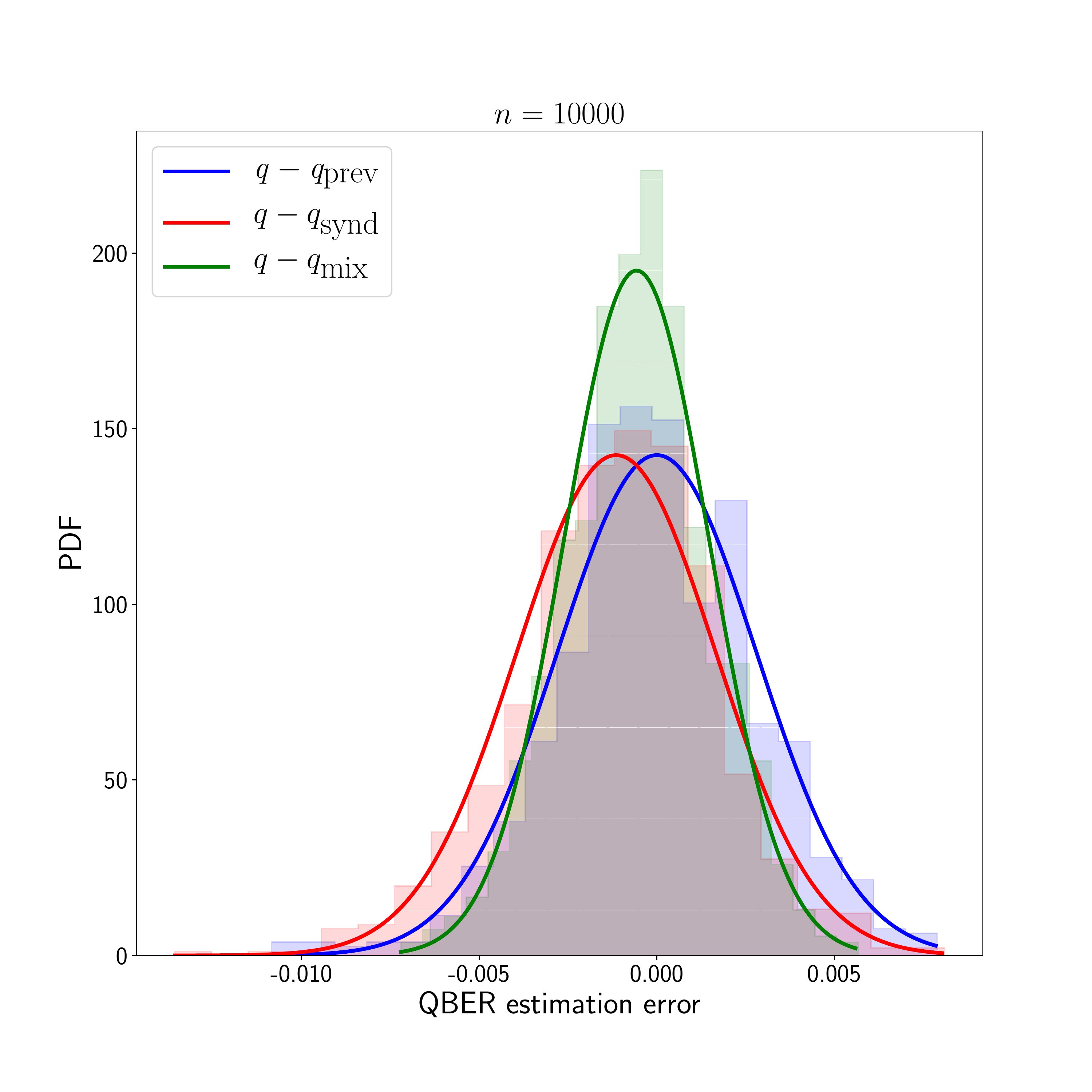}\\
		\includegraphics[height=0.35\linewidth]{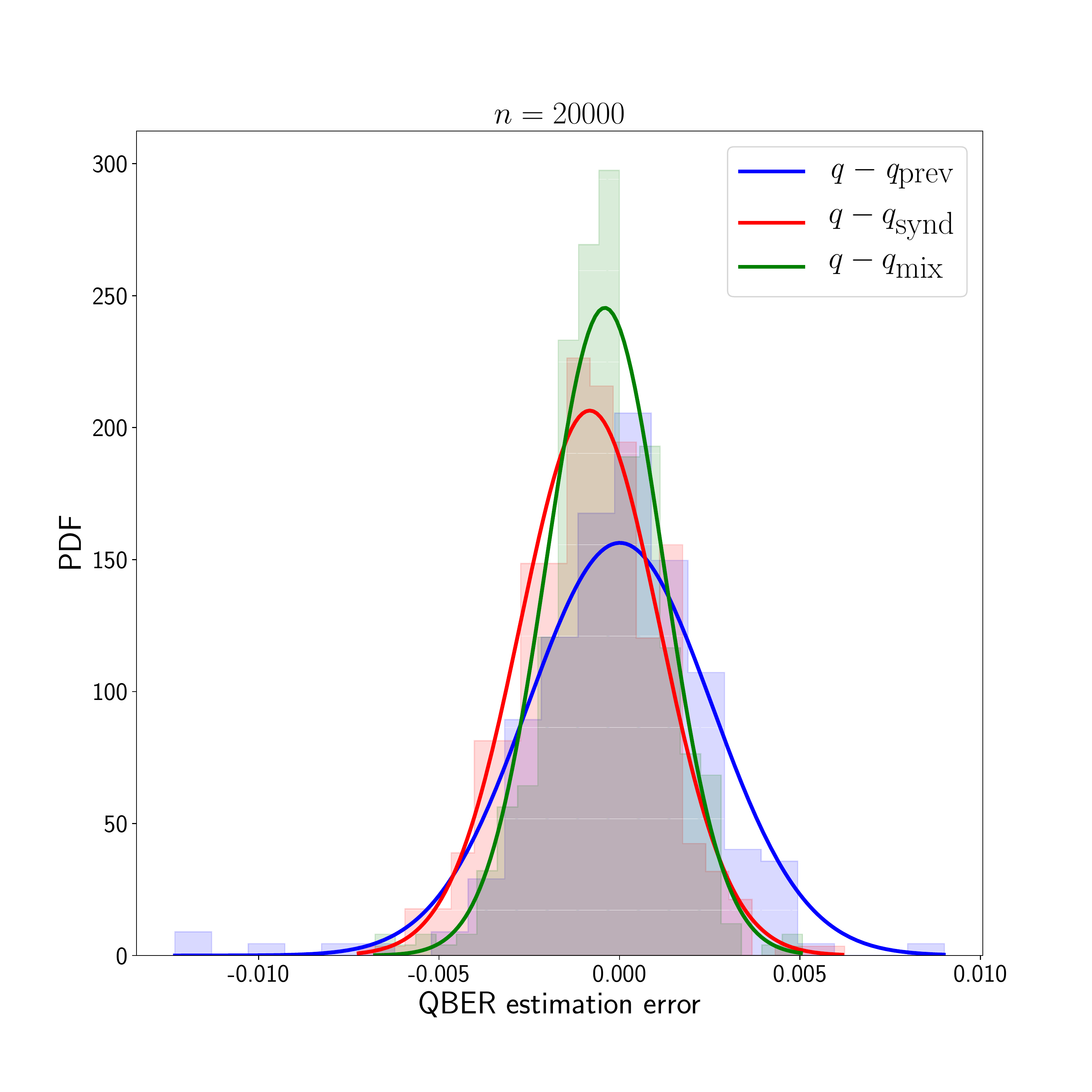} 
		\includegraphics[height=0.35\linewidth]{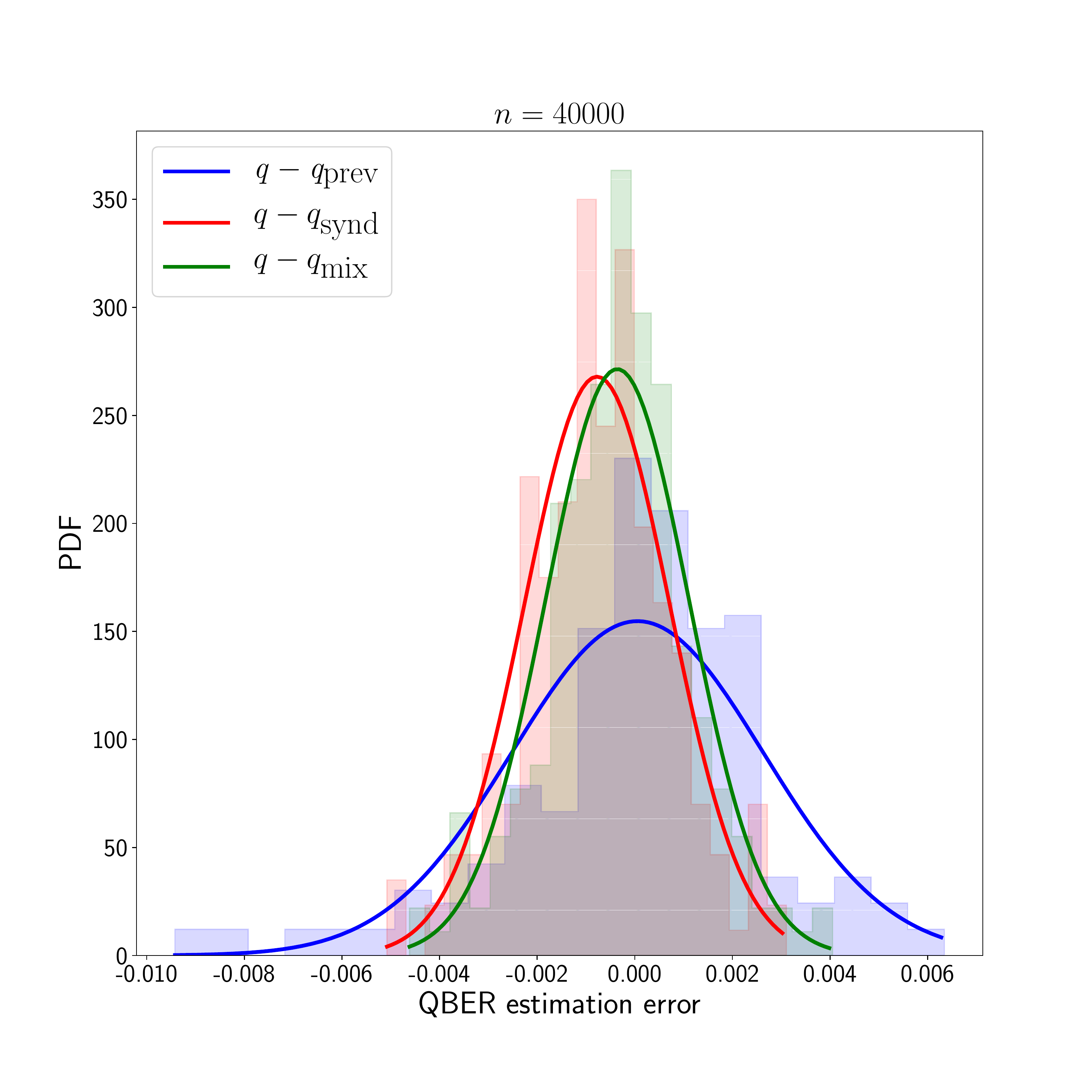}
		\caption{Probability density functions of $q-q_{\rm prev}$ (dashed line), $q-q_{\rm synd}$ (dotted line), and $q-q_{\rm mix}$ (solid line)  for LDPC codes with frame lengths $n = 4,000$ (a), 10,000 (b), 20,000 (c), 40,000 (d).}
		\label{fig:histograms}
	\end{figure*}
	
	Here we consider an implementation of our approach for a pair of sifted keys generated using the industrial QKD setup. 
	The setup is described in Ref.~\cite{Duplinskiy2018}.
	We employ four pools of LDPC codes with frame lengths $n = 4000, 10000, 20000, 40000$.
	Each pool consists of nine codes with rates $\mathcal{R}=\{0.5, 0.55, 0.6, \ldots 0.9\}$.
	The codes were generated using the improved edge growth algorithm~\cite{IPEG}, that was applied to the degree distribution polynomials from Ref.~\cite{LDPC2_poly}.
	To provide efficient fine-tuning of code rate with puncturing and shortening techniques we set $n_d = 0.05n$.
	With such the value of $n_d$ we can obtain an arbitrary code rate in range from $R=9/19\approx 0.47$ up to $R =18/19\approx 0.947$. 
	For the a priori QBER distribution, we use the function given by Eq.~(\ref{equ:sigmoid}) with $\alpha_1=\alpha_2=500$, $q_{\rm min}=0.01$ and $q_{\rm max}=0.08$.
	The value of $q_{\rm min}$ corresponds to minimal value of the QBER achievable on the employed QKD setup, while $q_{\rm max}$ is critical value of QBER which allows distillation of a secret key.
	
	In the demonstration presented below, we consider sifted keys of 4Mbit total length.
	The behaviour of the QBER calculated for block of $n=4000$ is shown in Fig.~\ref{fig:qber_to_time}.
	One can see that actual values of the QBER are indeed in the chosen ($q_{\tr{min}}, q_{\tr{max}}$) range.
	
	\begin{table*}
		\centering
		\begin{tabular}{|l|c|c|c|c|}
			\hline
			Frame length $n$ & 4000 & 10000 & 20000 & 40000 \\ \hline
			Number of sifted key bits $n-n_s-n_p$ & 3800 & 9500 & 19000 & 38000 \\ \hline
			\begin{tabular}[c]{@{}l@{}}Accuracy of the error estimation using \\ previous QBER $\sqrt{\overline{(q-q_{\rm prev})^2}}$\end{tabular} 
			& 0.0040 & 0.0028 & 0.0026  & 0.0026  \\ \hline
			\begin{tabular}[c]{@{}l@{}}Bias of the error estimation \\ using previous QBER   $\overline{q-q_{\rm prev}}$\end{tabular} 
			& $1.3\cdot10^{-6}$ & $5.4\cdot10^{-6}$ & $1.1\cdot 10^{-5}$  & $5.6\cdot 10^{-5}$  \\ \hline
			\begin{tabular}[c]{@{}l@{}}Accuracy of the error estimation \\ using syndrome  $\sqrt{\overline{(q-q_{\rm synd})^2}}$\end{tabular}
			& 0.0055 & 0.0030 & 0.0021 & 0.0017  \\ \hline
			\begin{tabular}[c]{@{}l@{}}Bias of the error estimation \\ using syndrome   $\overline{q-q_{\rm synd}}$\end{tabular} 
			& -0.0020 & -0.0011 & -0.00082  & -0.00077  \\ \hline
			\begin{tabular}[c]{@{}l@{}}Accuracy of the error estimation \\ in the mixed approach  $\sqrt{\overline{(q-q_{\rm mix})^2}}$\end{tabular}
			& 0.0034 & 0.0021 & 0.0017 & 0.0015  \\ \hline
			\begin{tabular}[c]{@{}l@{}}Bias of the error estimation \\ in the mixed approach   $\overline{q-q_{\rm mix}}$\end{tabular} 
			& -0.0010 & -0.00057 & -0.00041  & -0.00036  \\ \hline
		\end{tabular}
		\caption{Comparison results of the considered error estimation approaches for different frame lengths $n$.}\label{tab:approach_comparison}
	\end{table*}
	
	We then compare three approaches to the QBER estimation in sifted keys blocks of length $0.95n$ for each value of $n$.
	First, we consider a one using QBER value from the previous block ($q_{\rm prev}$) as an estimation of the QBER in the current block $q$.
	Second, we consider the approach described in Ref.~\ref{sec:QBER_estimator}. 
	To obtain a syndrome $q_{\rm synd}$ we choose a code from the codes pool and extend sifted key block of length $0.95n$ bits with $n_s$ shortened and $n_p$ punctured bits ($n_s+n_p=n_d$) such that the following approximate equality holds:
	\begin{equation}\label{eq:chooser}
	\frac{m-n_p}{(n-n_d)h(q_{\rm prev})} \approx 1.
	\end{equation}
	In most cases this equality can not be strictly fulfilled since $m$, $n_p$, and $n_s$ are integers.
	It should be noted that given the available rates $\mathcal{R}$ and total number of shortened and punctured bits $n_d=0.05n$ there is only one appropriate code rate from $\mathcal{R}$ that fulfills the condition~\eqref{eq:chooser}.
	Finally, we consider a mixed approach with the QBER estimation taken as the average of two aforementioned approaches: $q_{\rm mix}:=0.5(q_{\rm prev}+q_{\rm synd})$.
	
	To evaluate performance of the approaches we measure such performance metrics as bias and accuracy of the estimation, i.e. mean and root mean square of the $q - q_{\tr{est}}$ random variable corresponding to the error of an estimation approach. 
	The Value $q_{\tr{est}}$ is the QBER estimation obtained with one of the approaches $\left( q_{\tr{est}} \in \left \lbrace q_{\tr{prev}}, q_{\tr{synd}}, q_{\tr{mix}} \right \rbrace\right)$.
	
	The results of all three methods performance evaluation are given in Fig.~\ref{fig:histograms} and Table~\ref{tab:approach_comparison}.
	Fig.~\ref{fig:histograms} shows the histograms of $q - q_{\tr{est}}$  random variable values observed in the experiments. Table~\ref{tab:approach_comparison} contains measured values of bias and accuracy for different frame lengths.
	One can see that for small length codes ($n=4000$) the syndrome-based approach is not as precise as the one using previous value of QBER since the latter has the accuracy by 30~\% better.
	Nevertheless, for $n=10000$ performance of these two approaches becomes comparable and for $n=40000$ syndrome-based approach outperforms the first one.
	We also observe rather high bias of the syndrome-based approach. It turns out that the estimate of the error appears to typically higher than the actual error level.
	The estimate using previous QBER value does not suffer from such shortcoming.
	
	The most interesting result is that for all frame lengths the mixed approach provides the most accurate estimation.
	The advantage of the mixed approach comes from the fact that two first estimates are uncorrelated and the bias caused by syndrome-based estimation diminishes when taking into account the estimation using previous QBER.
	Thus, the average of two estimations turns out to be better than each of the estimations separately.
	This is an important observation of the present study.
	
	It should be noted that in $q_{\rm mix}$ we considered equal weighting factors (1/2) for $q_{\rm prev}$ and $q_{\rm synd}$.
	It seems that the accuracy of the mixed approach may be improve by adaptive changing of these factors according to the accuracy of two basic estimators. Also, we may add some constant to decrease the bias from syndrome-based approach.
	However, these improvements are beyond the scope of the current paper and are the directions of our further research. 
	
	\section{Conclusion}\label{sec:Conclusion}
	
	QKD systems are the prominent solution of guaranteeing secure data transmission even under the threat of quantum computer existence. 
	A precise estimation of QBER level at the stage of information reconciliation is crucial for the efficiency of QKD systems.
	In this paper, we have considered the problem of the accurate QBER estimation at the stage of information reconciliation of the QKD post-processing procedure. 
	We have proposed a novel method for the QBER estimation, which employs likelihood of a syndrome obtained from LDPC codeword.
	In contrast to previous works, our approach is relevant for  irregular LDPC codes and in the presence of both shortened and punctured bits as well. Moreover, we have provided a practical form for an a priori QBER distribution.
	
	We have applied our approach to sifted keys generated in the industrial QKD setup. 
	We have shown that combination of the proposed approach with the approach using error value from the previous round allows improving the accuracy of QBER estimation.
	We note that the use of proposed approach should be promising in different ways of using LDPC codes for information reconciliation within QKD post-processing.
	Particularly, it seems to be useful in rate-adaptive~\cite{Elkouss2010, Kronberg2017}, blind~\cite{LDPC_blind}, and symmetric-blind information reconciliation protocols~\cite{Kiktenko2017}.
	
	Finally, we would like to mention that  on-the-fly syndrome-based error estimation is especially promising in cases of highly unstable quantum channel which may undergo large sudden fluctuations in its properties.
	In this case, the accurate QBER estimation is able to provide a notable improvement in the secret key generation rate.
	
	\textbf{Acknowledgments}.
	The work was supported by the RFBR (Grant No. 18-37-00096).

\end{document}